\documentclass[12pt]{iopart}

\usepackage{graphicx}
\begin{document}

\title[Punch-through jets in $A+A$ collisions at RHIC/LHC]
{Punch-through jets in $A+A$ collisions at RHIC/LHC}

\author{Hanzhong Zhang$^1$, J.~F. Owens$^2$, Enke Wang$^1$ and X.-N. Wang$^3$}

\address{$^1$Institute of Particle Physics, Huazhong Normal University,
         Wuhan 430079, China; Key Laboratory of Quark \& Lepton
         Physics (HZNU), MOE,
         China}
\address{$^2$Physics Department, Florida State University, Tallahassee,
          FL 32306-4350, USA}
\address{$^3$Nuclear Science Division, Lawrence Berkeley Laboratory,
         Berkeley, CA 94720, USA}
\ead{zhanghz@iopp.ccnu.edu.cn}

\begin{abstract}
High $p_T$ single and dihadron production is studied within a NLO
pQCD parton model with jet quenching in high energy $A+A$ collisions
at the RHIC/LHC energy. A simultaneous $\chi^2$-fit to both single
and dihadron spectra can be achieved within a narrow range of energy
loss parameter. Punch-through jets are found to result in the
dihadron suppression factor slightly more sensitive to medium than
the single hadron suppression factor at RHIC. Such jets at LHC are
found to dominate high $p_T$ dihadron production and the resulting
dihadron spectra are more sensitive to the initial parton
distribution functions than the single hadron spectra.
\end{abstract}


A strongly coupled quark gluon plasma (sQGP) is now believed to
exist in high energy nucleus-nucleus collisions with a large volume
and a long life time. Such dense matter can be probed by the
quenching\cite{JQ-82-92} of high transverse momentum $p_T$ partonic
jets which will lose a significant amount of their energy via
induced gluon radiations when propagating through the dense matter.
The energy loss is predicted to lead to strong suppression of both
single- and correlated away-side dihadron spectra at high $p_T$
~\cite{xnw04,zoww07}, consistent with experimental
findings~\cite{phenix-star-single,star03-06}.

\begin{figure}[thb]
\hspace{1.3in}
\includegraphics[width=0.6\linewidth]{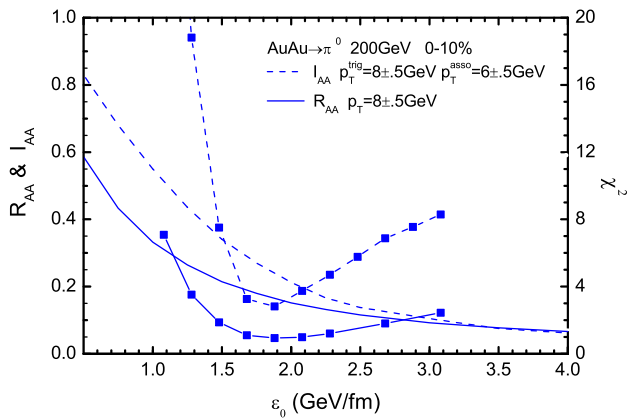}
\vskip -.3in \caption{\label{fig:chi2} The suppression factors for
single ($R_{AA}$) and dihadron ($I_{AA}$) spectra at fixed
transverse momentum as functions of the initial energy loss
parameter $\epsilon_0$. Also shown are $\chi^2$'s (curves with
filled squares) in fitting experimental data on single ($p_T=4 - 20
$ GeV/$c$) and away-side spectra ($p_T^{\rm trig}=8-15$ GeV,
$z_T=0.45-0.95$) in central $Au+Au$ collisions at $\sqrt{s}=200$
GeV.}
\end{figure}

\begin{figure}[tb]
\hspace{.2in}
\includegraphics[width=0.4\linewidth]{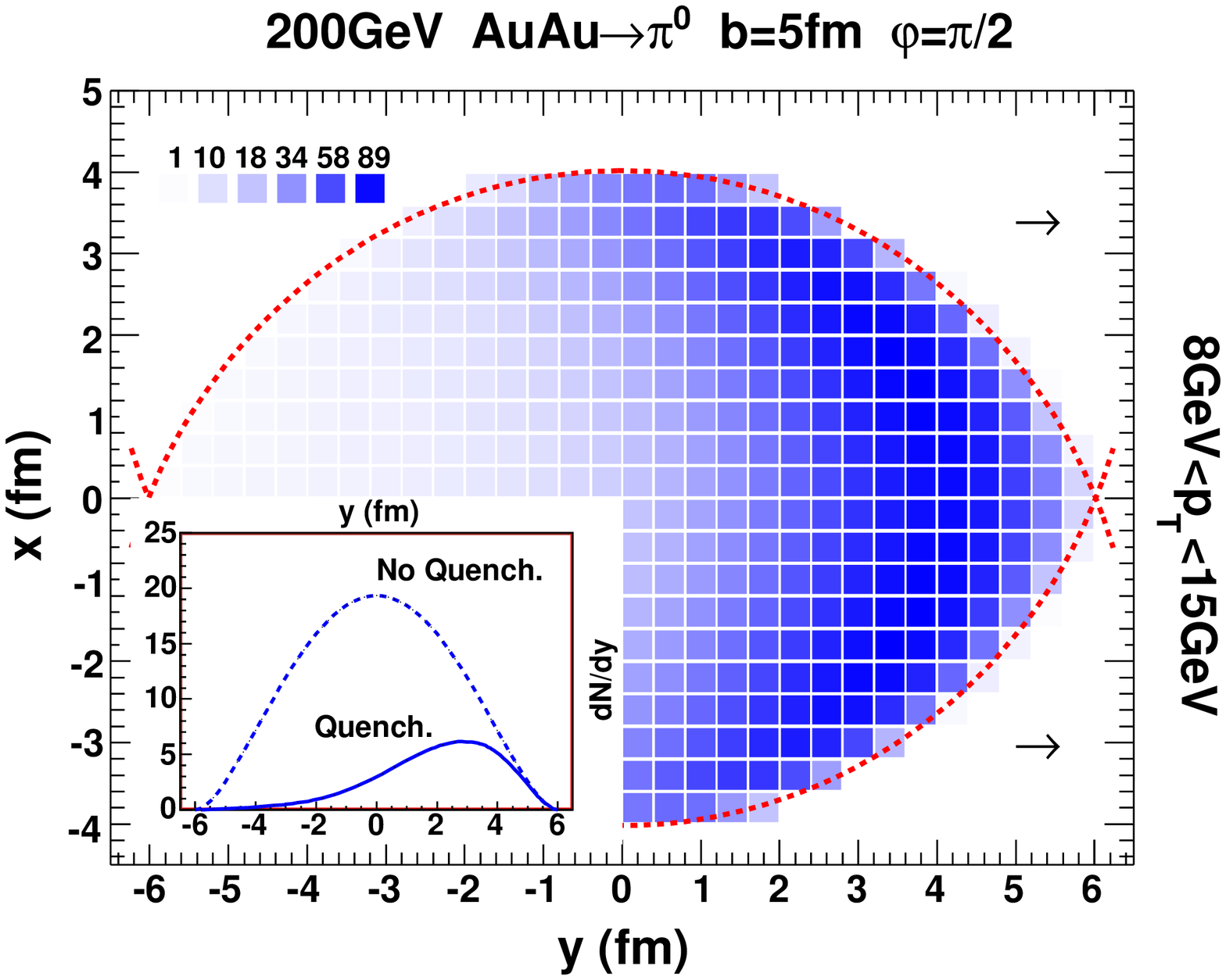}
\vskip -2.0in\hspace{3.3in}
\includegraphics[width=0.4\linewidth]{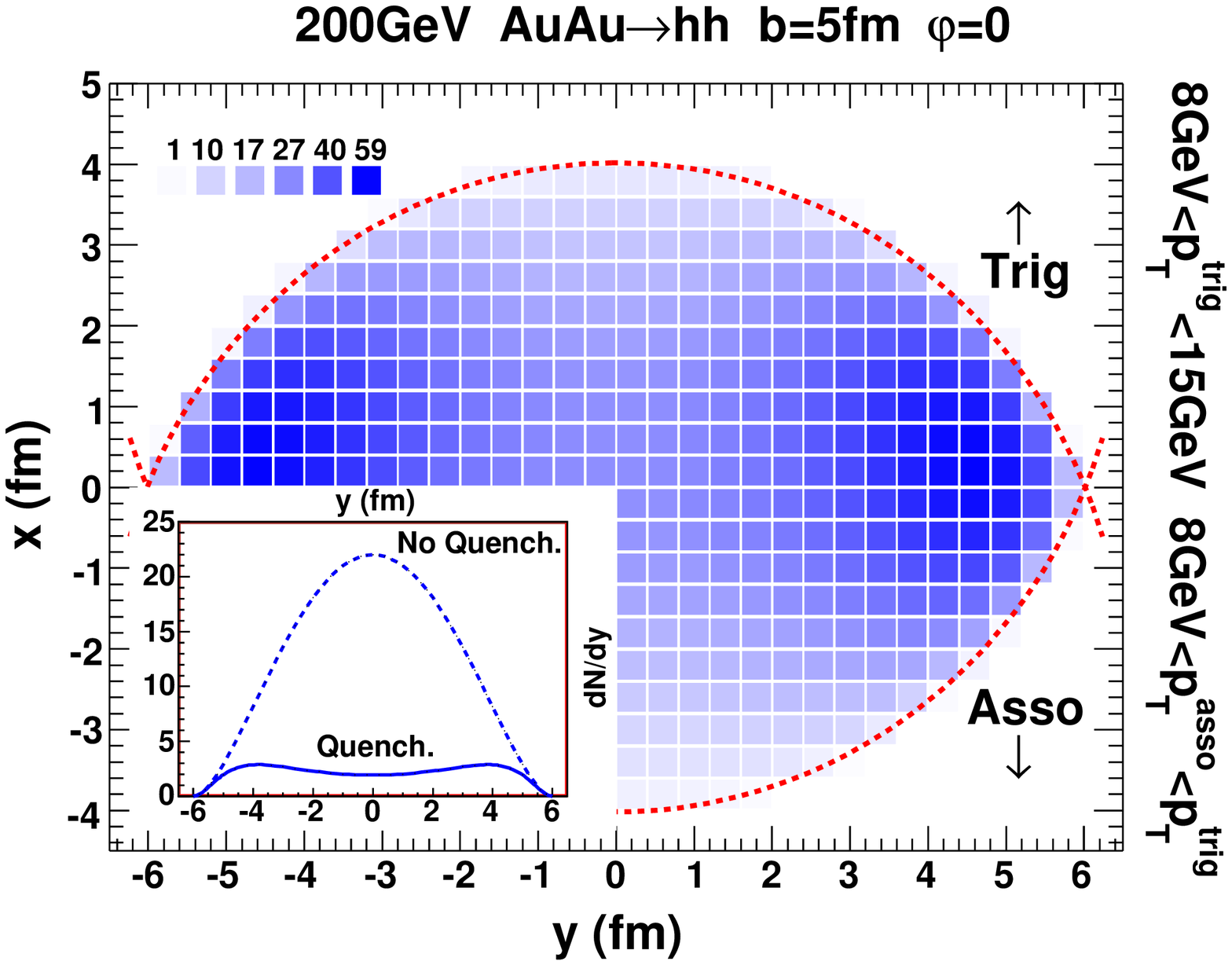}
\vskip -.15in \caption{\label{fig:cont-rhic} Spatial transverse
distribution (arbitrary normalization) of the initial parton
production points that contribute to the single and dihadron along a
given direction at RHIC. The insert is the same distribution
projected onto the $y$-axis.}
\end{figure}

The suppression factor of the leading hadrons from jet fragmentation
will depend on the total parton energy loss which in turn is related
to the jet propagation path weighted with the gluon density $\rho_g$
along the propagation path~\cite{xnw04}. A simultaneous fit to the
single and dihadron data constrains the energy loss parameter within
a narrow range: $\epsilon_0=1.6-2.1$ GeV/fm~\cite{zoww07}. The fact
that both $\chi^2$'s in Figure~(\ref{fig:chi2}) reach their minima
in the same range for two different measurements provides convincing
evidence for the jet quenching description. At the RHIC energy, the
high $p_T$ dihadrons are found to come not only from jet pairs close
and tangential to the surface of the dense matter but also from
punch-through jets originating from the center of the system while
the high $p_T$ single hadrons are only dominated by jets emitted
perpendicularly near the surface of the overlap in
Figure~(\ref{fig:cont-rhic}). Consequently, the dihadron spectra is
slightly more sensitive to the initial gluon density than the single
spectra.

\begin{figure}[tb]
\hspace{.2in}
\includegraphics[width=0.4\linewidth]{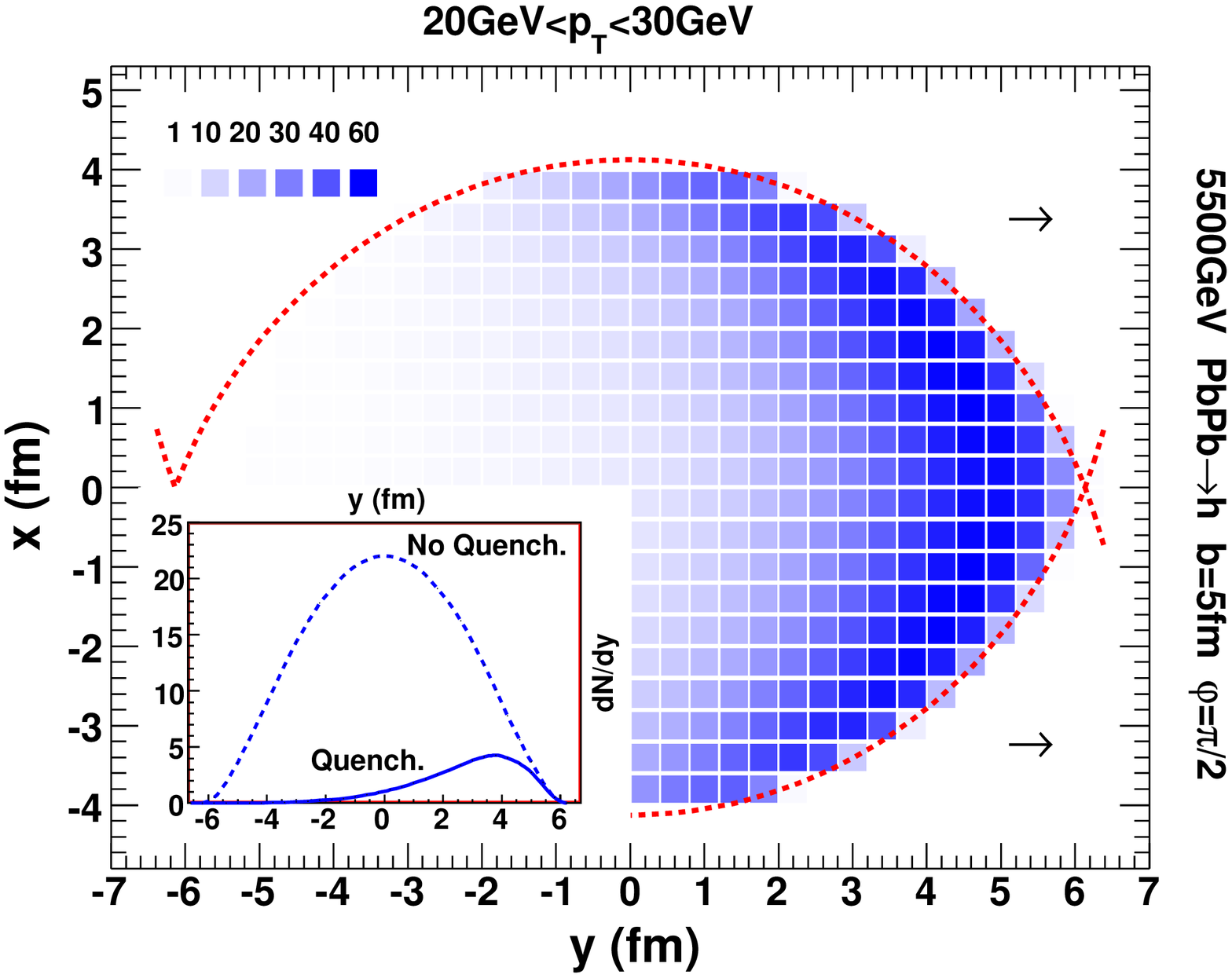}
\vskip -2.05in\hspace{3.3in}
\includegraphics[width=0.4\linewidth]{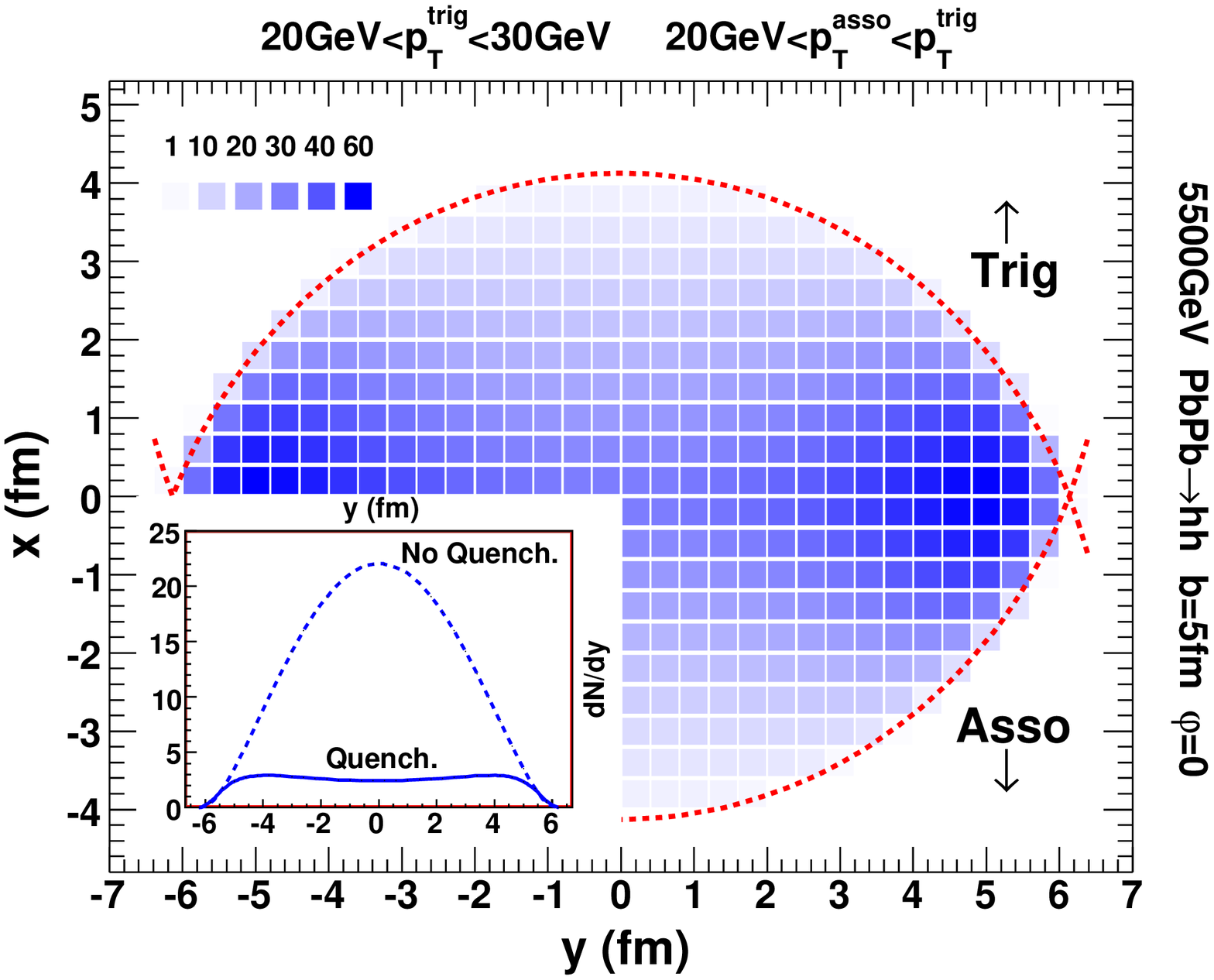}
\vskip -.0in

\hspace{.2in}
\includegraphics[width=0.4\linewidth]{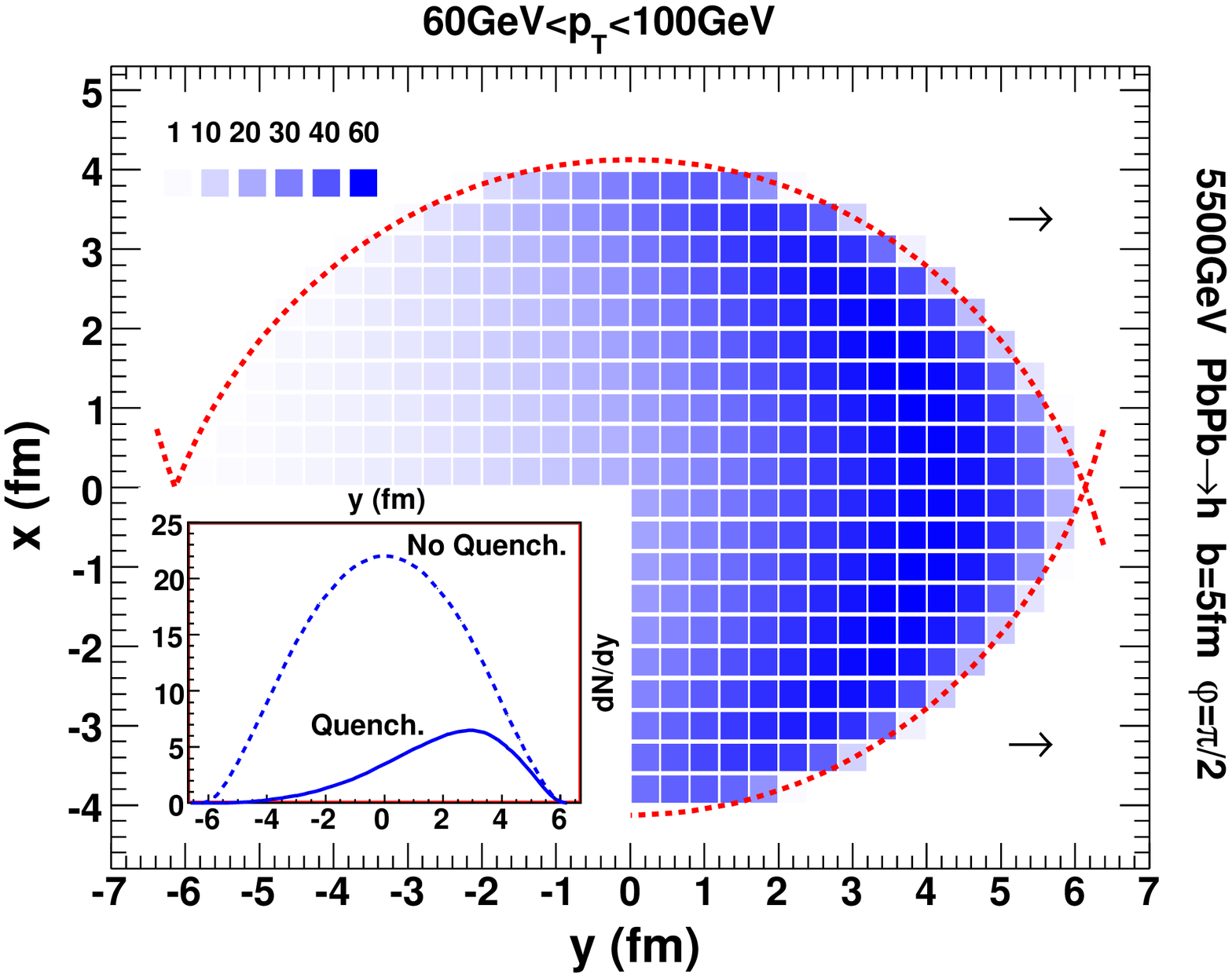}
\vskip -2.05in\hspace{3.3in}
\includegraphics[width=0.4\linewidth]{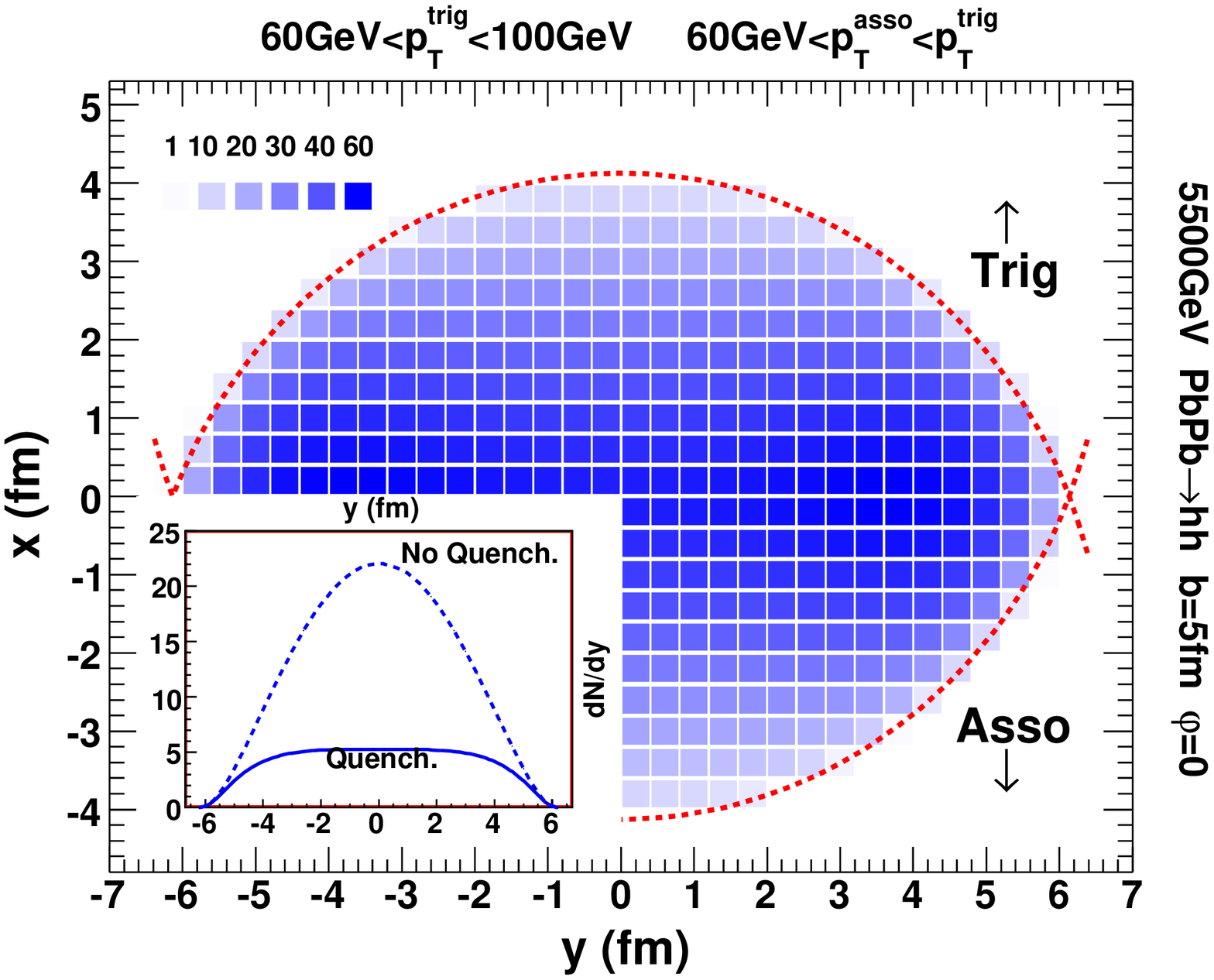}
\vskip .0in

\hspace{.2in}
\includegraphics[width=0.4\linewidth]{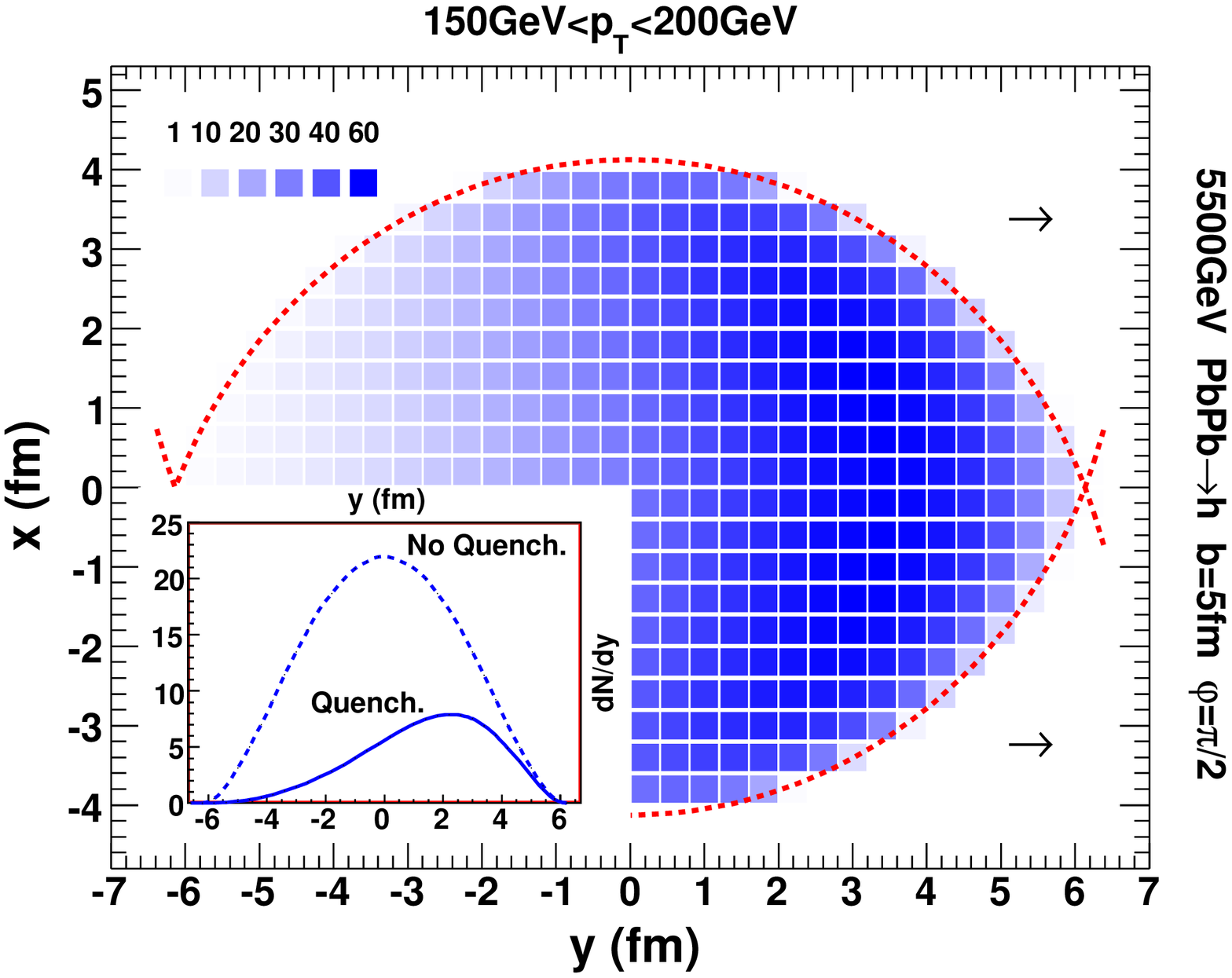}
\vskip -2.05in\hspace{3.3in}
\includegraphics[width=0.4\linewidth]{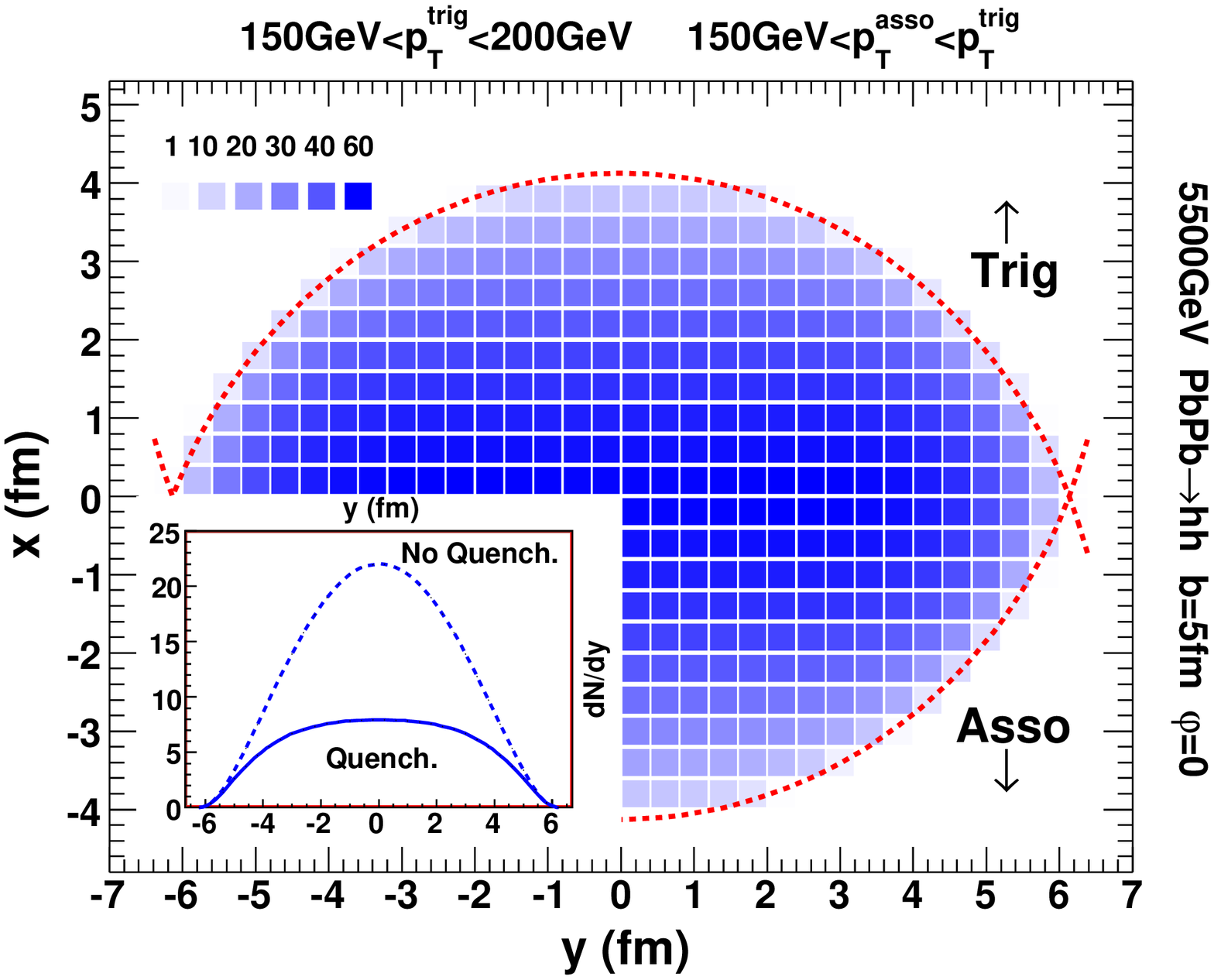}
\vskip -.15in \caption{\label{fig:cont-lhc} The same as
Figure~(\ref{fig:cont-rhic}) but at the LHC energy.}
\end{figure}

Punch-through jets are significant for dihadron production not only
at the RHIC energy but also at the LHC energy. Shown in
Figure~(\ref{fig:cont-lhc}) are the spatial transverse distributions
of the initial parton production points that contribute to the final
high $p_T$ single and dihadron at the LHC energy. The jets
contributing to the single hadron at LHC still have a surface
emission bias similar to the RHIC case in
Figure~(\ref{fig:cont-rhic}). However, the fraction of dihadron
yield from punch-through jets is found to increase with the
transverse momenta of dihadron at the LHC energy. Such difference in
the geometry of the single hadron and dihadron production will
result in some interesting phenomena.

\begin{figure}[tb]
\hspace{.7in}
\includegraphics[width=.85\linewidth]{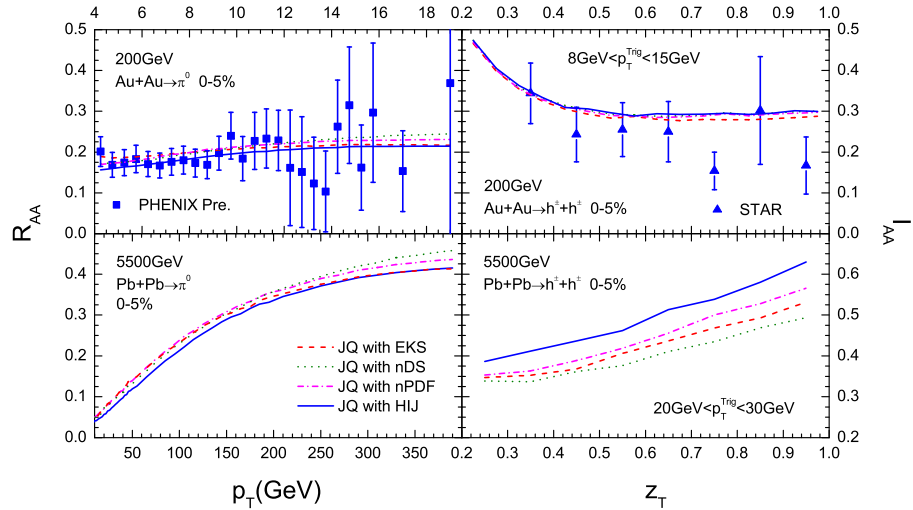}
\vskip -.2in \caption{\label{fig:RaaIaa-shad-2} Four kinds of
shadowing effects checked in the single and dihadron suppression
factors at the RHIC/LHC energy. The data are from
Ref.~\cite{akiba,phenix07-pi0}.}
\end{figure}

Shown in Figure~(\ref{fig:RaaIaa-shad-2}) are the single and
dihadron suppression factors in central $Au+Au$ collisions with 4
kinds of shadowing parameterizations, EKS98~\cite{eks98},
nDS~\cite{nDS}, nPDF~\cite{nPDF}, Hijing~\cite{hijing}. Among the 4
sets of suppression factors, $R_{AA}/I_{AA}$ at RHIC/LHC, only the
dihadron suppression factor $I_{AA}$ at LHC is found to be sensitive
to different shadowing parameterizations. Initial partons
participating in strong interaction in the central region should be
associated with stronger shadowing effects than those initial
partons in the outer layer of the system. Because of surface
emission bias, the single hadron suppression factor is insensitive
to shadowing at both the RHIC and LHC energy. Punch-through jets
originate from the central region where the shadowing is stronger.
Such jets at RHIC are greatly suppressed, so the dihadron
suppression factor is also insensitive to the shadowing at RHIC.
However, the punch-through jets at LHC are found to dominate
dihadron spectra, so the dihadron suppression factor is sensitive to
shadowing at LHC.

In summary, high $p_T$ single and dihadron productions are studied
within a NLO pQCD parton model with jet quenching in $A+A$
collisions at the RHIC/LHC energy. A simultaneous $\chi^2$-fit to
both single and dihadron spectra can be achieved within their minima
in the same narrow range of energy loss parameter for two different
measurements. This fact provides a convincing evidence for jet
quenching description. Punch-through jets are found to result in the
dihadron suppression factor slightly more sensitive to medium than
the single suppression factor at RHIC. Such jets at LHC are found to
dominate dihadron production with increasing dihadron transverse
momentum, and therefore dihadron suppression is found to be more
sensitive to different shadowing parameterizations of initial parton
distributions at LHC.

This work was supported by DOE under Contracts No. DE-AC02-05CH11231
and No. DE-FG02-97IR40122, by MOE of China under Projects No.
IRT0624, No. NCET-04-0744 and No. SRFDP-20040511005, and by NSFC of
China under Projects No. 10440420018, No. 10475031 and No. 10635020.

\end{document}